\documentclass[aps,prl,showpacs,reprint,superscriptaddress]{revtex4-1}

\usepackage{amsmath, amssymb}
\usepackage{graphicx}
\usepackage{bm}

\newcommand{\blr}[1]{ \left( #1 \right) }
\newcommand{\bclr}[1]{ \left[ #1 \right] }
\renewcommand{\vec}[1]{ \boldsymbol{ #1 } }
\renewcommand{\vr}[0]{ \vec{r} }
\newcommand{\vq}[0]{ \vec{q} }
\newcommand{\ind}[1]{ \int\!\!\mathrm{d}#1 \; }
\newcommand{\cH}[0]{ \hat{ \mathcal{H} } }
\newcommand{\cT}[0]{ \hat{ \mathcal{T} } }
\newcommand{\cV}[0]{ \hat{ \mathcal{V} } }
\newcommand{\cB}[0]{ \hat{ \mathcal{B} } }
\newcommand{\cW}[0]{ \hat{ \mathcal{W} } }
\newcommand{\fder}[2]{ \frac{\delta#1}{\delta#2} }
\renewcommand{\S}[1]{ \sin\!\blr{ #1 } }
\newcommand{\C}[1]{ \cos\!\blr{ #1 } }
\newcommand{\abs}[1]{ |#1|}
\newcommand{\nn}[0]{ \nonumber \\}
\newcommand{\PFr}[0]{ \hat{ \Phi } \! \blr{\vr} }
\newcommand{\PFdr}[0]{ \hat{ \Phi }^{ \dagger } \! \blr{\vr} }
\newcommand{\U}[0]{ \mathcal{U} }

\newcommand{\uFFdr}[0]{ \hat{ \phi }_{\uparrow}^{ \dagger } \!\blr{\vr} }

\newcommand{\dFFdr}[0]{ \hat{ \phi }_{\downarrow}^{ \dagger } \!\blr{\vr} }

\begin{document}

\title{Transverse spin gradient functional for non-collinear \\ spin density functional theory}

\author{F.~G.~Eich}
\email[]{feich@mpi-halle.mpg.de}
\affiliation{Max-Planck-Institut f{\"u}r Mikrostrukturphysik, Weinberg 2, D-06120 Halle, Germany}
\affiliation{Institut f{\"u}r Theoretische Physik, Freie Universit{\"a}t Berlin, Arnimallee 14, D-14195 Berlin, Germany}
\author{E.~K.~U.~Gross}
\affiliation{Max-Planck-Institut f{\"u}r Mikrostrukturphysik, Weinberg 2, D-06120 Halle, Germany}

\date{\today}

\begin{abstract}
  We present a novel functional for spin density functional theory aiming at the description
  of non-collinear magnetic structures. The construction of the functional employs the spin-spiral-wave state
  of the uniform electron gas as reference system. We show that the functional depends on transverse
  gradients of the spin magnetization, i.e.~, in contrast to the widely used local spin density
  approximation, the functional is sensitive to local changes of the direction of the spin magnetization.
  As a consequence the exchange-correlation magnetic field is not parallel to the spin magnetization
  and a local spin-torque is present in the ground state of the Kohn-Sham system. As a proof-of-principle we
  apply the functional to a Chromium mono-layer in the non-collinear ${120^{\circ}}$-N{\'e}el state.
\end{abstract}


\maketitle

Since the discovery of the giant magnetoresistance\cite{Fert:88,*Grunberg:89} the field of spintronics \cite{ZuticDasSarma:04} 
plays an important role in the everlasting goal to miniaturize devices for data storage and manipulation.
For instance, the coupling of orbital and spin degrees of freedom is used to move magnetic
domain walls in so-called ``racetrack'' memory devices \cite{Parkin:2008} via a charge current.
Similarly, spin polarized currents can switch the magnetic state of spin-valves by means of
the so-called spin-transfer torque.\cite{RalphStiles:08} Whenever spin-orbit coupling is present, there is no
global spin-quantization axis and the spin magnetization becomes \emph{non-collinear}.
A specific example of non-collinear magnetic structures on the nano scale are skyrmions \cite{Skyrme:62}, i.e.~, topological twists in the
spin magnetization, which recently have been observed in magnetic solids \cite{Muehlbauer:09} and magnetic surfaces.\cite{HeinzeBlugel:11} 
Even within a single atom non-collinear magnetism is present.\cite{NordstroemSingh:96,*EschrigServedio:99}

Density functional theory (DFT) \cite{DreizlerGross:90} is presently the most widely used approach
to determine the electronic structure of large molecules and solids. Shortly after the original formulation
by Hohenberg and Kohn \cite{HohenbergKohn:64} in terms of the electronic density ${n\blr{\vr}}$ alone,
the theory was extended to include also the spin magnetization ${\vec{m}\!\blr{\vr}}$ as fundamental variable.\cite{BarthHedin:72}
Spin density functional theory (SDFT) applies to Hamiltonians of the form
\begin{equation}
  \cH = \cT + \cV + \cB + \cW , \label{SDFT_H}
\end{equation}
where in addition to the kinetic energy ${\cT}$, the potential energy ${\cV}$ and the
electron-electron interaction energy ${\cW}$ a contribution, ${\cB}$,
due to an external magnetic field ${\vec{B}\!\blr{\vr}}$ is considered, i.e.~,
\begin{equation}
  \cB = - \mu_{\mathrm{B}} \ind{^3r} \vec{B}\!\blr{\vr} \cdot \hat{\vec{m}}\!\blr{\vr} . \label{B_m}
\end{equation}
The operator ${\hat{\vec{m}}\!\blr{\vr}= \PFdr \vec{\sigma} \PFr}$, representing the spin magnetization, is defined in
terms of the spinor field ${\PFdr = \begin{pmatrix} \uFFdr & \dFFdr \end{pmatrix}}$ 
and the vector of Pauli matrices ${\vec{\sigma}}$. 

An immediate application of spin density functional theory (SDFT) is to find the configuration of the spin magnetization which is lowest in energy and hence
the most stable. Furthermore it is possible to shape the spin magnetization via an external magnetic field. Comparing
energies for different magnetic configurations (or equivalently different external magnetic fields) one can map out the energy landscape for a
given material. A specific example is to constrain the spin magnetization to rotate in space with a given wave vector ${\vec{q}}$ in order to compute
the magnon dispersion by ${\epsilon_{mag}\!\blr{\vec{q}}=E\!\blr{\vec{q}}-E\!\blr{0}}$ (frozen magnon approach).

As always in DFTs the success of the theory hinges on the availability of accurate and physically sound
approximations to the exchange-correlation (${xc}$) energy ${E_{xc}\!\bclr{n,\vec{m}}}$ - a functional of ${n\!\blr{\vr}}$ and ${\vec{m}\!\blr{\vr}}$
in the case of SDFT. The functional derivative of ${E_{xc}\!\bclr{n,\vec{m}}}$ w.r.t.~the density (spin magnetization) yields the so-called ${xc}$ potential 
${v_{xc}\!\blr{\vr}}$ (${xc}$ magnetic field  ${\vec{B}_{xc}\!\blr{\vr}}$). These ${xc}$ potentials describe the effect of exchange and correlation
in the Kohn-Sham (KS) system \cite{KohnSham:65}, an effective system of non-interacting electrons, exposed to the potential
${v_{s}\!\blr{\vr}=v\!\blr{\vr}+v_{\mathrm{H}}\!\blr{\vr}+v_{xc}\!\blr{\vr}}$ and magnetic field ${\vec{B}_{s}\!\blr{\vr}=\vec{B}\!\blr{\vr}+\vec{B}_{xc}\!\blr{\vr}}$, 
which reproduces the density and spin magnetization of the interacting system.

The simplest approximation in the framework of DFT is the local density approximation (LDA) \cite{KohnSham:65}, which
determines the ${xc}$ energy of the non-uniform systems by treating it locally as a uniform electron gas.
Including the spin magnetization this idea is readily generalized yielding for the ${xc}$ energy
\begin{equation}
  E_{xc}^{\mathrm{LSDA}}\!\bclr{n,\vec{m}} = \ind{^3r} n\!\blr{\vr} \varepsilon_{xc}^{\mathrm{unif}}\!\blr{n\!\blr{\vr},m\!\blr{\vr}} , \label{LSDA}
\end{equation}
the so-called local spin density approximation (LSDA), with ${m\!\blr{\vr}}$ being the magnitude of ${\vec{m}\!\blr{\vr}}$ and
${\varepsilon_{xc}^{\mathrm{unif}}}$ the ${xc}$ energy of a spin-polarized uniform electron gas (UEG).
For collinear magnetism, i.e.~, ${\vec{m}\!\blr{\vr}}$ pointing in the same direction everywhere in space, a plethora of functionals
was derived (cf.~Ref.~\onlinecite{RappoportBurke:09}) improving over the \emph{collinear} LSDA, however, much
less is known about constructing functionals for \emph{non-collinear} magnetism, where the direction of ${\vec{m}\!\blr{\vr}}$ is allowed to
vary freely in space. In fact, most applications of non-collinear SDFT up-to-date are based on the idea, pioneered by K{\"u}bler et al.~\cite{KueblerWilliams:88},
to apply collinear functionals to non-collinear systems by evaluating the functional in a \emph{local} reference frame with the local ${z}$-axis
determined by the direction of ${\vec{m}\!\blr{\vr}}$. The LSDA, defined in Eq.~\eqref{LSDA}, employs a local reference frame intrinsically which
can be seen by evaluating the corresponding ${xc}$ magnetic field
\begin{equation}
  \vec{B}_{xc}^{\mathrm{LSDA}}\!\blr{\vr} = - \fder{E_{xc}^{\mathrm{LSDA}}\!\bclr{n,\vec{m}}}{\mu_{\mathrm{B}} \vec{m}\!\blr{\vr}} =
  - \frac{\partial \varepsilon_{xc}^{\mathrm{unif}}}{\partial m} \frac{n\!\blr{\vr} \vec{m}\!\blr{\vr}}{\mu_{\mathrm{B}}m\!\blr{\vr}} . \label{Bxc_LSDA}
\end{equation}
By construction ${\vec{B}_{xc}^{\mathrm{LSDA}}}$ is always aligned with ${\vec{m}}$. The same is true for generalized gradient approximations (GGAs)
employing the aforementioned rotation to a local reference frame. 
In recent years attempts were made to extend GGAs and meta-GGAs to non-collinear systems
without invoking a local reference frame
in order to produce a ${\vec{B}_{xc}\!\blr{\vr}}$ which is non-collinear w.r.t.~${\vec{m}\!\blr{\vr}}$.\cite{PeraltaFrisch:07,*ScalmaniFrisch:12} 
Since collinear functionals are usually formulated in terms of ${n_{\uparrow}\!\blr{\vr}}$
and ${n_{\downarrow}\!\blr{\vr}}$ (as opposed to ${n\!\blr{\vr}}$ and ${m\!\blr{\vr}}$) and gradients thereof, these approaches
require a prescription mapping the gradient of ${\vec{m}\!\blr{\vr}}$ (a ${3\!\times\!3}$-matrix for non-collinear systems)
to gradients of ${n_{\uparrow}\!\blr{\vr}}$ and ${n_{\downarrow}\!\blr{\vr}}$. 
Sharma et al.~demonstrated that orbital functionals yield in general a ${\vec{B}_{xc}}$ which is non-collinear w.r.t.~${\vec{m}}$.\cite{SharmaGross:07}
Another approach was to consider the variations of the direction of ${\vec{m}\!\blr{\vr}}$ perturbatively.\cite{KatsnelsonAntropov:03,CapelleGyorffy:03}
Capelle and Oliveira proposed a \emph{non-local} DFT approach \cite{CapelleOliveira:00a,CapelleOliveira:00b}, 
in close analogy to the DFT for superconductors.\cite{OliveiraKohn:88}

In this letter we show that the very idea of the LSDA can be extended in a non-perturbative way to yield a new functional
for SDFT depending on transverse gradients. This means that the ${xc}$ functional depends on spatial variations of the direction
of ${\vec{m}}$ and, as a consequence, the ${xc}$ magnetic field exerts a local torque on the spin magnetization.
This local torque is important for the ab-initio description of spin dynamics. \cite{CapelleGyoerffy:01} 

In the LSDA the spin polarized UEG is chosen as reference system to determine the local ${xc}$ energy. Note that the LSDA
does not employ the ground-state ${xc}$ energy of the UEG, but instead the minimal ${xc}$ energy of the UEG
under the \emph{constraint} that its spin magnetization is ${m_{0}}$. Usually one imposes the constraint of a fixed
spin magnetization via a uniform magnetic field. 
The new functional is based on the idea to consider a reference system with a non-collinear spin magnetization.
In close analogy to the LSDA the
local ${xc}$ energy is determined from the UEG constrained to be in the so-called spin-spiral-wave (SSW) state.\cite{Overhauser:62, *GiulianiVignaleSDW:05} 
The SSW state of the UEG is characterized by a constant density ${n_0}$ and a spin magnetization of the form
\begin{equation}
  \vec{m}_{0}\!\blr{\vr} = m_{0} \begin{pmatrix} s \C{\vq\!\cdot\!\vr} \\ s \S{\vq\!\cdot\!\vr} \\ \sqrt{1-s^2} \end{pmatrix} , \label{m_SSW}
\end{equation}
with ${s=\S{\theta}}$ and ${\theta}$ is the azimuthal angle between the rotating part (in the ${x}$-${y}$ plane)
and the constant part (parallel to ${z}$-axis). Similar to the case of the uniformly polarized UEG the constraint of 
a spin-spiral magnetization is imposed via a local external magnetic field that itself has a spiral structure.
\footnote{This is shown explicitly for a non-interacting electron gas in the supplemental material.}
The ${xc}$ energy of the SSW UEG depends on four parameters: ${n_0}$, ${m_0}$, ${s}$ and ${q=\abs{\vq}}$. As we will see below it is possible 
to define \emph{local} ${s\!\blr{\vr}}$ and ${q\!\blr{\vr}}$ in terms of transverse gradients of ${\vec{m}\!\blr{\vr}}$ which
leads to the definition of the SSW functional
\begin{equation}
  E_{xc}^{\mathrm{SSW}}\!\bclr{n,\vec{m}} = \ind{^3r} n\!\blr{\vr} \varepsilon_{xc}^{\mathrm{SSW}}\!\blr{n\!\blr{\vr},m\!\blr{\vr},s\!\blr{\vr},q\!\blr{\vr}} , \label{SSW}
\end{equation}
where ${\varepsilon_{xc}^{\mathrm{SSW}}\!\blr{n,m,s,q}}$ is the minimal ${xc}$ energy of the UEG under the \emph{constraint} that it is in the SSW state
specified by ${n}$, ${m}$, ${s}$ and ${q}$. It is important to realize that the LSDA is included
in this definition in the limits ${s \to 0}$ or ${q \to 0}$, i.e.~, 
$\varepsilon_{xc}^{\mathrm{SSW}}\!\blr{n,m,s,q=0}=\varepsilon_{xc}^{\mathrm{SSW}}\!\blr{n,m,s=0,q}=\varepsilon_{xc}^{\mathrm{SSW}}\!\blr{n,m,s=0,q=0}=\varepsilon_{xc}^{\mathrm{unif}}\!\blr{n,m}$. 
This can be emphasized by rewriting the SSW functional as
\begin{align}
  E_{xc}^{\mathrm{SSW}}\!\bclr{n,\vec{m}} & = \ind{^3r} n\!\blr{\vr} \varepsilon_{xc}^{\mathrm{unif}}\!\blr{n\!\blr{\vr},m\!\blr{\vr}} \nn
  & \times \blr{1+S_{xc}\!\blr{n\!\blr{\vr},m\!\blr{\vr},s\!\blr{\vr},q\!\blr{\vr}}} , \label{SSW_SGE}
\end{align}
where we have introduced the spin gradient enhancement (SGE)
\begin{equation}
  S_{xc}\!\blr{n,m,s,q} = \frac{\varepsilon_{xc}^{\mathrm{SSW}}\!\blr{n,m,s,q}-\varepsilon_{xc}^{\mathrm{unif}}\!\blr{n,m}}{\varepsilon_{xc}^{\mathrm{unif}}\!\blr{n,m}} . \label{SGE}
\end{equation}

Before we discuss the explicit form of the local ${s\!\blr{\vr}}$ and ${q\!\blr{\vr}}$ we briefly discuss \emph{global}, i.e.~, spatially
independent, rotations of the internal (spin) space. These rotations correspond to transforming ${\PFr \to \U \PFr}$, 
where ${\U}$ is an element of ${\mathrm{SU}\!\blr{2}}$ (a rotation of the internal [spin] degree of freedom). Note that \emph{spatial} vectors, e.g.~the
(charge) current ${\vec{\jmath}\!\blr{\vr}}$, are invariant under such internal rotations whereas \emph{spin} vectors as ${\vec{m}\!\blr{\vr}}$
transform as ${\vec{m}\!\blr{\vr} \to \underline{R} \vec{m}\!\blr{\vr}}$, with ${\underline{R}}$ being the ${3\!\times\!3}$ rotation matrix
corresponding to ${\U}$. Since the kinetic energy ${\cT}$ and the interaction energy ${\cW}$ are invariant under global
rotations of the internal space it follows that ${E_{xc}\!\bclr{n, \vec{m}} = E_{xc}\!\bclr{n, \underline{R}\vec{m}}}$. Considering
infinitesimal spin rotations one obtains the so-called zero-torque theorem
\begin{equation}
  0 = \ind{^3r} \vec{m}\!\blr{\vr} \times \vec{B}_{xc}\!\blr{\vr} , \label{ZTT}
\end{equation}
which was first derived by Capelle et al.~via the equation of motion for the spin magnetization.\cite{CapelleGyoerffy:01} It states
that ${\vec{B}_{xc}}$ \emph{cannot} exert a net torque on the whole system.

A simple rule to follow in order to ensure that explicit functionals for SDFT obey the zero-torque theorem is to write ${E_{xc}\!\bclr{n,\vec{m}}}$ in terms
of proper scalars, i.e.~, spin indices have to be contracted with spin indices and spatial indices with spatial indices. 
This implies that the determination of the local ${xc}$ energy
in terms of \emph{strictly} local densities is exhausted by ${n\!\blr{\vr}}$ and ${m\!\blr{\vr}}$. Hence the local ${s\!\blr{\vr}}$ and
${q\!\blr{\vr}}$ have to be determined from properly contracted gradients of ${\vec{m}\!\blr{\vr}}$.

Let us first look at ${D\!\blr{\vr}=\abs{\nabla\!\otimes\!\vec{m}\!\blr{\vr}}^2}$, which corresponds to the \emph{total} first order change
of ${\vec{m}\!\blr{\vr}}$. It can be split into a \emph{longitudinal} contribution ${D_{\mathrm{L}}\!\blr{\vr}}$ and a \emph{transverse}
contribution ${D_{\mathrm{T}}\!\blr{\vr}}$, i.e.~,
\begin{align}
  D\!\blr{\vr} & = \frac{1}{m^2\!\blr{\vr}} \blr{ D_{\mathrm{L}}\!\blr{\vr} + D_{\mathrm{T}}\!\blr{\vr} } , \label{D_DL_DT} \\
  D_{\mathrm{L}}\!\blr{\vr} & = \abs{\vec{m}\!\blr{\vr} \!\cdot\! \blr{\nabla \!\otimes\! \vec{m}\!\blr{\vr}}}^2 , \label{DL} \\
  D_{\mathrm{T}}\!\blr{\vr} & = \abs{\vec{m}\!\blr{\vr} \!\times\! \blr{\nabla \!\otimes\! \vec{m}\!\blr{\vr}}}^2 , \label{DT}
\end{align}
where the meaning of longitudinal and transverse is defined by the local direction of ${\vec{m}\!\blr{\vr}}$.
We use ``${\otimes}$'' to emphasize that the gradient of the magnetization is a tensor. 
The scalar and the cross product in Eqs.\ \eqref{DL} and \eqref{DT} act on the components of ${\vec{m}}$.
\footnote{Note that ``${\abs{\ldots}^2}$'' always implies a contraction of the remaining indices, e.g.\ , ${\abs{\nabla\otimes\vec{m}}^2=\blr{\partial_{i}m_{j}}\blr{\partial_{i}m_{j}}}$
where a summation of repeated indices is implied.}
For the SSW UEG the two contributions are ${D_{\mathrm{L}}^{\mathrm{SSW}}=0}$ and ${D_{\mathrm{T}}^{\mathrm{SSW}}=m_{0}^4 s^2 q^2}$. 
Both contributions are constant in space for the SSW UEG and hence play a similar role as the density ${n_{0}}$ and the magnitude of the spin magnetization ${m_{0}}$,
i.e.~, they locally characterize the state.
${D_{\mathrm{L}}^{\mathrm{SSW}}}$ vanishes because the spin magnetization in the SSW UEG only rotates (the magnitude ${m}$ is constant). ${D_{\mathrm{T}}}$ does not
vanish but it only determines the combination ${s q}$. 

Accordingly we look at the second order variation ${d\!\blr{\vr}=\abs{\nabla^2 \vec{m}\!\blr{\vr}}^2}$.
Again, it can be analyzed w.r.t.~longitudinal and transverse contributions
\begin{align}
  d\!\blr{\vr} & = \frac{1}{m^2\!\blr{\vr}} \blr{ d_{\mathrm{L}}\!\blr{\vr} + d_{\mathrm{T}}\!\blr{\vr} } , \label{d_dL_dT} \\
  d_{\mathrm{L}}\!\blr{\vr} & = \abs{\vec{m}\!\blr{\vr} \!\cdot\! \blr{\nabla^2 \vec{m}\!\blr{\vr}}}^2 , \label{dL} \\
  d_{\mathrm{T}}\!\blr{\vr} & = \abs{\vec{m}\!\blr{\vr} \!\times\! \blr{\nabla^2 \vec{m}\!\blr{\vr}}}^2 . \label{dT}
\end{align}
For our reference system this yields ${d_{\mathrm{L}}^{\mathrm{SSW}}=m_{0}^4 s^4 q^4}$ and ${d_{\mathrm{T}}^{\mathrm{SSW}}=m_{0}^4 \blr{1-s^2} s^2 q^4}$. 
The change of ${\vec{m}\!\blr{\vr}}$ to first order is perpendicular to ${\vec{m}\!\blr{\vr}}$, but to second order ${\vec{m}\!\blr{\vr}}$
also changes in the direction of ${\vec{m}\!\blr{\vr}}$ which explains why ${d_{\mathrm{L}}^{\mathrm{SSW}}}$ does not vanish for the SSW UEG.
However, we see that ${d_{\mathrm{L}}^{\mathrm{SSW}}}$ provides \emph{the same} information as
${D_{\mathrm{T}}^{\mathrm{SSW}}}$, meaning, ${s q}$ to some power. Adopting the strategy
that we obtain the characteristic parameters for the local ${xc}$ energy choosing the order of derivatives
as low as possible, ${s\!\blr{\vr}}$ and ${q\!\blr{\vr}}$ are given by 
\begin{align}
  s\!\blr{\vr} & = \sqrt{\frac{D_{\mathrm{T}}^2\!\blr{\vr}}{D_{\mathrm{T}}^2\!\blr{\vr} + m^4\!\blr{\vr} d_{\mathrm{T}}\!\blr{\vr}}} , \label{s_local} \\
  q\!\blr{\vr} & = \sqrt{\frac{D_{\mathrm{T}}^2\!\blr{\vr} + m^4\!\blr{\vr} d_{\mathrm{T}}\!\blr{\vr}}{m^4\!\blr{\vr} D_{\mathrm{T}}\!\blr{\vr}}} . \label{q_local}
\end{align}
This completes the definition of the SSW functional Eq.~\eqref{SSW} or equivalently the SGE to the LSDA Eq.~\eqref{SSW_SGE}.

By definition (c.f.~Eq.~\eqref{s_local}) the local ${s\!\blr{\vr}}$ is between ${\bclr{0,1}}$ in accordance with being the sine of an azimuthal angle.
Furthermore we have the following hierarchy in the dependence of the SGE, Eq.~\eqref{SSW_SGE}, on the transverse gradients: 
i) If ${D_{\mathrm{T}}\!\blr{\vr}=0}$, the SGE correction is zero.
ii) If ${D_{\mathrm{T}}\!\blr{\vr} \neq 0}$ and ${d_{\mathrm{T}}\!\blr{\vr}=0}$,
the SGE correction is obtained from a \emph{planar} SSW (${s=1}$). 
iii) If both transverse gradients are non-zero the SGE correction is obtained from a general SSW.

We proceed by evaluating the ${xc}$ magnetic field from the SSW functional,
\begin{equation}
  \vec{B}_{xc}^{\mathrm{SSW}}\!\blr{\vr} = - \fder{E_{xc}\!\bclr{n,\vec{m}}}{\mu_{\mathrm{B}}\vec{m}\!\blr{\vr}}
  = \vec{B}_{xc}^{m} + \vec{B}_{xc}^{D_{\mathrm{T}}} + \vec{B}_{xc}^{d_{\mathrm{T}}} , \label{Bxc_SSW}
\end{equation}
where we split ${\vec{B}_{xc}^{\mathrm{SSW}}\!\blr{\vr}}$ into contributions coming from the dependence of ${\varepsilon_{xc}^{\mathrm{SSW}}}$ on
${m}$, ${D_{\mathrm{T}}}$ and ${d_{\mathrm{T}}}$, respectively. The explicit evaluation of ${\vec{B}_{xc}^{\mathrm{SSW}}\!\blr{\vr}}$
is straight-forward but rather lengthy.
Here we will show the energetic content in the KS system, i.e.~,
\begin{align}
  & E_{\vec{B}_{xc}}^{\mathrm{KS}}  = - \mu_{\mathrm{B}} \ind{^3r} \vec{m}\!\blr{\vr} \!\cdot\! \vec{B}_{xc}\!\blr{\vr} \nn
  & = 2 \ind{^3r} n\!\blr{\vr} \blr{\partial_{m^2}\varepsilon_{xc}^{\mathrm{SSW}}} \abs{\vec{m}\!\blr{\vr}}^2 \label{EKS_B_m} \\
  & + 4 \ind{^3r} n\!\blr{\vr} \blr{\partial_{D_{\mathrm{T}}}\varepsilon_{xc}^{\mathrm{SSW}}} \abs{\vec{m}\!\blr{\vr} \!\times\! \blr{\nabla \!\otimes\! \vec{m}\!\blr{\vr}}}^2 \label{EKS_B_D} \\
  & + 4 \ind{^3r} n\!\blr{\vr} \blr{\partial_{d_{\mathrm{T}}}\varepsilon_{xc}^{\mathrm{SSW}}}  \abs{\vec{m}\!\blr{\vr} \!\times\! \blr{\nabla^2 \vec{m}\!\blr{\vr}}}^2 . \label{EKS_B_d}
\end{align}
The first term (Eq.~\eqref{EKS_B_m}) is already present in the LSDA, whereas the other two terms (Eqs.~\eqref{EKS_B_D},\eqref{EKS_B_d}) arise due to the
inclusion of the SGE.
The zero-torque theorem, Eq.~\eqref{ZTT}, is fulfilled by construction, however the new terms in ${\vec{B}_{xc}\!\blr{\vr}}$
are non-collinear w.r.t.~${\vec{m}\!\blr{\vr}}$, i.e.~, they provide a local torque.

The final step for a practical implementation of the SSW functional is the determination of the SGE from the SSW UEG.
We have evaluated ${S_{xc}\!\blr{n,m,s,q}}$ using the random-phase approximation (RPA) for the SSW UEG. It is important to stress
that we approximate the ${S_{xc}}$ with the RPA and \emph{not} ${\varepsilon_{xc}^{\mathrm{SSW}}}$. In this way the SSW functional
reduces to the LSDA parameterized using the Monte Carlo reference data.\cite{CeperleyAlder:80,*PerdewWang:92}
From ${\sim 65000}$ data points in the four-dimensional domain of ${S_{xc}}$ we have constructed a polynomial fit for ${S_{xc}}$.

\begin{figure}[htp]
  \includegraphics{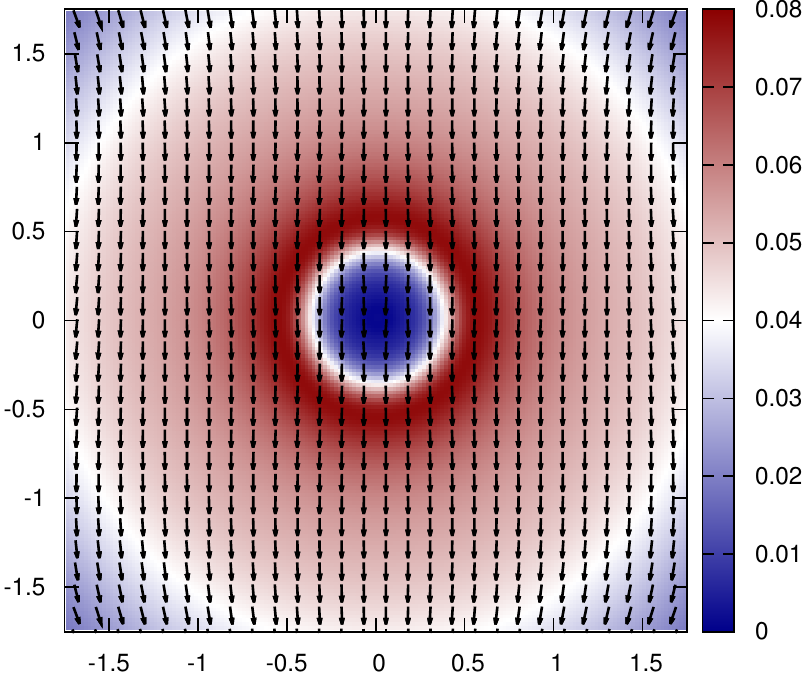}
  \caption{
    (Color online). Magnitude (color coded) and direction (arrows) of ${\vec{B}_{xc}\!\blr{\vr}}$ around an atom for the Cr mono-layer using the LSDA. 
    \label{FIG_BXC_LSDA}
  }
\end{figure}

\begin{figure}[htp]
  \includegraphics{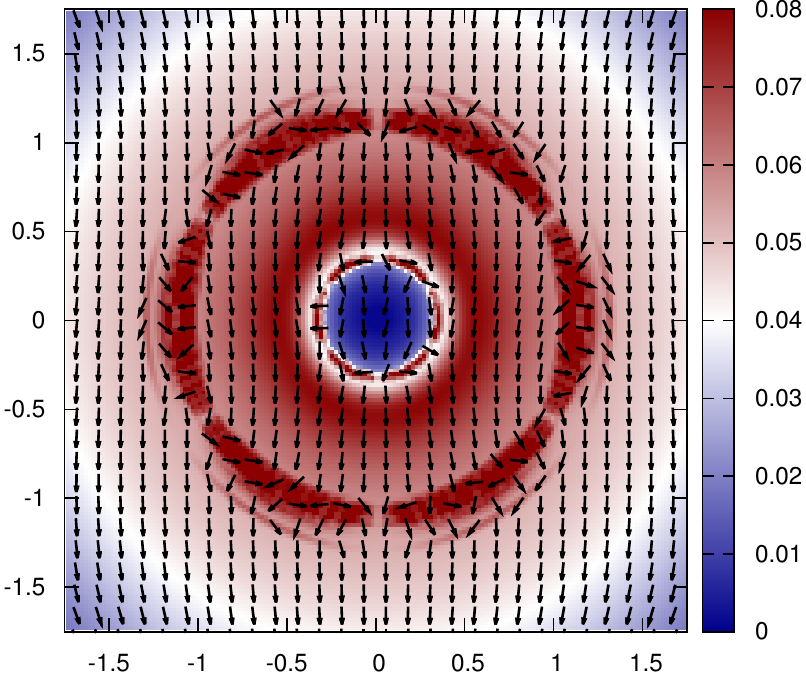}
  \caption{
    (Color online). Same as FIG.~\ref{FIG_BXC_LSDA} for the SSW functional (LSDA+SGE).
    Note the richer structure of ${\vec{B}_{xc}\!\blr{\vr}}$ close to the Cr nucleus.
    \label{FIG_BXC_SGE}}
\end{figure}

\begin{figure}[hbp]
  \includegraphics{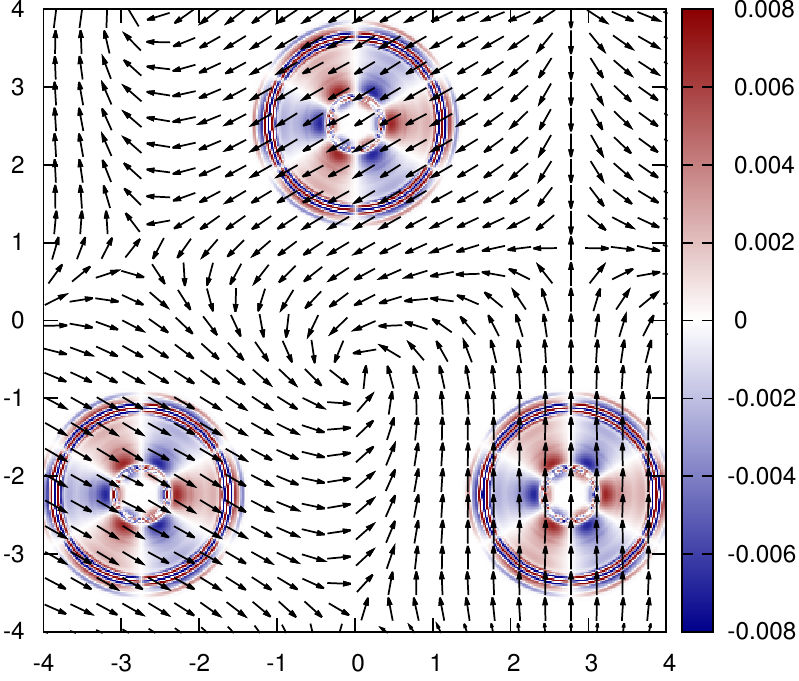}
  \caption{
    (Color online). The ${z}$-component of ${\vec{m}\!\blr{\vr}\!\times\!\vec{B}_{xc}\!\blr{\vr}}$ (color coded) around
    Cr atoms in the mono-layer computed using the SGE functional. The arrows show the direction of the spin magnetization.
    \label{FIG_MCBXC}}
\end{figure}

As a first application we have implemented the SSW functional in the ELK code \cite{ELK} in order to investigate
the Chromium mono-layer in the ${120^{\circ}}$-N{\'e}el state. In FIGs.~\ref{FIG_BXC_LSDA} and \ref{FIG_BXC_SGE}
we plot the magnitude and direction of ${\vec{B}_{xc}}$ in order to illustrate the qualitative difference between
the LSDA and the SSW functional. While the local spin magnetizations ${\vec{m}\!\blr{\vr}}$ are similar for the
LSDA and the SSW functional, ${\vec{B}_{xc}}$ obtained via the SGE exhibits much more structure compared to the
LSDA ${\vec{B}_{xc}}$. As a result the local torque does not vanish and a ground-state
spin current is present in the KS system. \footnote{This can be seen from the equation of motion for the spin magnetization (cf.~Ref.~\onlinecite{CapelleGyoerffy:01}).}
The local torque which is completely missed by the usual LSDA is shown explicitly in FIG.~\ref{FIG_MCBXC} for the non-collinear the ${120^{\circ}}$-N{\'e}el state.
The \emph{global} zero-torque theorem (cf.~Eq.~\eqref{ZTT}) may be inferred from the pattern
of negative (blue) and positive (red) local torques around the nuclei. Since the SSW functional
is not restricted to small ${q\!\blr{\vr}}$ it accounts for the intra-atomic non-collinearity.

In conclusion we have proposed a novel functional for SDFT depending on the first and second order transverse gradients
of ${\vec{m}\!\blr{\vr}}$. We emphasize that this functional is formulated in terms of an enhancement
to the LSDA. In particular this means that the correction vanishes in the case of a collinear system. 
The construction of the new functional parallels closely the original formulation of the LSDA.
On one hand this means that the system is \emph{locally} treated as a uniform electron gas
in the SSW state, which may appear as a rather crude approximation. On the other hand 
the success of DFT may be attributed, to some extent, to the fact that the LDA already represents a reasonable
approximation even for strongly inhomogeneous systems.
GGAs are also corrections to the LSDA, hence it is conceivable to employ the two corrections
simultaneously. Since GGAs are constructed having collinear systems in mind, one may argue that
the longitudinal gradients ${D_{\mathrm{L}}\!\blr{\vr}}$, ${d_{\mathrm{L}}\!\blr{\vr}}$ should enter in the GGA part. 
We expect that the SGE will improve the ab-initio description
of materials exhibiting a non-collinear magnetic structure.

While the corrections to the part of the ${\vec{B}_{xc}\!\blr{\vr}}$ parallel to ${\vec{m}\!\blr{\vr}}$
will adjust the energetics, the perpendicular part of ${\vec{B}_{xc}\!\blr{\vr}}$ describes the ${xc}$ corrections
to the spin current, which in turn is crucial for ab-initio spin dynamics. We expect that the
functional presented in this letter will pave the road to a better description of domain wall motion and spin wave
propagation from first principles in the framework of time-dependent SDFT. 

In both aforementioned scenarios it is important to have a numerically accessible functional which, given currently available
computing facilities, implies the use of semi-local functionals. We have demonstrated that non-collinearity can be included by
a generalization of the reference system employed in the LSDA and hence the numerical accessibility of the LSDA
is retained in the SSW functional making it the ideal candidate for large scale quantum simulations.

\begin{acknowledgments}
  This study was partially supported by the Deutsche
  Forschungsgemeinschaft within the SFB 762, and by the
  European Commission within the FP7 CRONOS project (ID 280879).
  F. G. E. acknowledges useful discussions with Giovanni Vignale and Kay Dewhurst.
\end{acknowledgments}

\bibliography{biblio}

\end{document}